\documentclass{article}
\usepackage{spconf,amsmath,graphicx,booktabs,tikz,hyperref,fancyhdr}
\usetikzlibrary{backgrounds}
\usetikzlibrary{positioning}
\usetikzlibrary{calc}
\usetikzlibrary{shapes.multipart}
\usetikzlibrary{arrows.meta}
\usetikzlibrary{decorations.pathreplacing}
\usetikzlibrary{fit}

\fancypagestyle{firstpage}{%
\fancyhf{}
\fancyfoot[C]{\copyright2020 IEEE. To be published in the IEEE 2020 International Conference on Acoustics, Speech, and Signal Processing (ICASSP 2020), scheduled for 4-8 May 2018 in Barcelona, Spain.}

}

\title{Phonetic Feedback for Speech Enhancement \\ With and Without Parallel Speech Data}
\name{Peter Plantinga, Deblin Bagchi, Eric Fosler-Lussier}
\address{The Ohio State University}
\begin{document}
%
\maketitle
\thispagestyle{firstpage}
\begin{abstract}
While deep learning systems have gained significant ground in speech enhancement research, these systems have yet to make use of the full potential of deep learning systems to provide high-level feedback. In particular, phonetic feedback is rare in speech enhancement research even though it includes valuable top-down information. We use the technique of mimic loss to provide phonetic feedback to an off-the-shelf enhancement system, and find gains in objective intelligibility scores on CHiME-4 data. This technique takes a frozen acoustic model trained on clean speech to provide valuable feedback to the enhancement model, even in the case where no parallel speech data is available. Our work is one of the first to show intelligibility improvement for neural enhancement systems without parallel speech data, and we show phonetic feedback can improve a state-of-the-art neural enhancement system trained with parallel speech data.
\end{abstract}
\begin{keywords}
speech intelligibility, phonetic feedback, mimic loss, CHiME-4, parallel data
\end{keywords}
\section{Introduction}
\label{sec:intro}

Typical speech enhancement techniques focus on local criteria for improving speech intelligibility and quality. Time-frequency prediction techniques use local spectral quality estimates as an objective function; time domain methods directly predict clean output with a potential spectral quality metric \cite{pandey2019new}. Such techniques have been extremely successful in predicting a speech denoising function, but also require parallel clean and noisy speech for training. The trained systems {\em implicitly} learn the phonetic patterns of the speech signal in the coordinated output of time-domain or time-frequency units. However, our hypothesis is that directly providing phonetic feedback can be a powerful additional signal for speech enhancement. For example, many local metrics will be more attuned to high-energy regions of speech, but not all phones of a language carry equal energy in production (compare /v/ to /ae/).

Our proxy for phonetic intelligibility is a frozen automatic speech recognition (ASR) acoustic model trained on clean speech; the loss functions we incorporate into training encourage the speech enhancement system to produce output that is interpretable to a fixed acoustic model as clean speech, by making the output of the acoustic model mimic its behavior under clean speech. This \textit{mimic loss} \cite{bagchi2018spectral} provides key linguistic insights to the enhancement model about what a recognizable phoneme looks like.

When no parallel data is available, but transcripts are available, a loss is easily computed against hard senone labels and backpropagated to the enhancement model trained from scratch. Since the clean acoustic model is frozen, the only way for the enhancement model to improve the loss is to make a signal that is more recognizable to the acoustic model. The improvement by this model demonstrates the power of phonetic feedback; very few neural enhancement techniques until now have been able to achieve improvements without parallel data.

When parallel data is available, mimic loss works by comparing the outputs of the acoustic model on clean speech with the outputs of the acoustic model on denoised speech. This is a more informative loss than the loss against hard senone labels, and is complimentary to local losses. We show that mimic loss can be applied to an off-the-shelf enhancement system and gives an improvement in intelligibility scores. Our technique is agnostic to the enhancement system as long as it is differentiably trainable.

Mimic loss has previously improved performance on robust ASR tasks \cite{bagchi2018spectral}, but has not yet demonstrated success at enhancement metrics, and has not been used in a non-parallel setting. We seek to demonstrate these advantages here:


\begin{enumerate}
    \item We show that using hard targets in the mimic loss framework leads to improvements in objective intelligibility metrics when no parallel data is available.
    \item We show that when parallel data is available, training the state-of-the-art method with mimic loss improves objective intelligibility metrics.
\end{enumerate}

\begin{figure*}[t!]
\tikzstyle{model} = [rectangle, draw, minimum height=4em, minimum width=5em, align=center]
\tikzstyle{circ} = [circle, draw, align=center]
\centering
\begin{tikzpicture}
	\node[align=center] (noisy) at (0,0) {Noisy \\ speech};
    \node[model, right=1em of noisy] (aecnn) {AECNN \\ denoiser};
    \node[right=1em of aecnn,align=center,minimum height=3em,minimum width=5em] (denoised) {Denoised \\ speech};
    \node[above=4em of denoised,align=center, minimum height=3em, minimum width=5em] (clean) {Clean \\ speech};
    
    \node[circ, fill=gray!30, right=1em of clean] (stft1) {STFT};
    \node[circ, fill=gray!30, right=1em of denoised] (stft2) {STFT};
    
    \node[align=center, right=1em of stft1] (clean_mag) {Clean \\ spectral mag.};
    \node[align=center, right=1em of stft2] (denoised_mag) {Denoised \\ spectral mag.};
    
    \node[model, right=1em of clean_mag, fill=gray!30] (am1) {Acoustic \\ Model};
    \node[right=1em of am1,align=center,minimum height=3em, minimum width=5em] (soft_labels) {Soft \\ labels};
    
    \node[model, right=1em of denoised_mag, fill=gray!30] (am2) {Acoustic \\ model};
    \node[right=1em of am2,align=center,minimum height=3em,minimum width=5em] (out) {Network \\ outputs};
    
    \node[circ, right=1em of out, fill=gray!30] (softmax) {Softmax};
    \node[right=1em of softmax] (post) {Posterior};
    
    \draw[-stealth,thick] (noisy) -- (aecnn);
    \draw[-stealth,thick] (aecnn) -- (denoised);
    \draw[stealth-stealth,thick] (denoised) -- node[left,pos=0.5] (l1_1) {\large $L_1$} (clean);
    
    \draw[-stealth,thick] (clean) -- (stft1);
    \draw[-stealth,thick] (stft1) -- (clean_mag);
    \draw[-stealth,thick] (denoised) -- (stft2);
    \draw[-stealth,thick] (stft2) -- (denoised_mag);
    
    \draw[stealth-stealth,thick] (denoised_mag) -- node[left] (l1_2) {\large $L_1$} (clean_mag);
    
    \draw[-stealth,thick] (clean_mag) -- (am1);
    \draw[-stealth,thick] (am1) -- (soft_labels);
    \draw[-stealth,thick] (denoised_mag) -- (am2);
    \draw[-stealth,thick] (am2) -- (out);
    
    \draw[stealth-stealth,thick] (out) -- node[left] (l1_3) {\large $L_1$} (soft_labels);
    
    \node[above=1em of soft_labels, color=red] (par) {Parallel};
    \node[draw,dashed,red,fit=(par) (clean) (l1_1)] (rect1) {};
    
    \draw[-stealth,thick] (out) -- (softmax);
    \draw[-stealth,thick] (softmax) -- (post);
    \node[align=center] (hard) at (soft_labels -| post) {Hard senone \\ labels};
    \draw[stealth-stealth,thick] (post) -- node[left] (lce) {\large $L_{CE}$} (hard);

    \node[above=1em of hard, color = red] (non) {Non-parallel};
    \node[draw,dashed,red,fit=(non) (hard) (lce)] (rect2) {};

\end{tikzpicture}
\caption{Operations are listed inside shapes, the circles are operations that are not parameterized, the rectangles represent parameterized operations. The gray operations are not trained, meaning the loss is backpropagated without any updates until the front-end denoiser is reached.}
\label{fig:systemdesc}
\end{figure*}
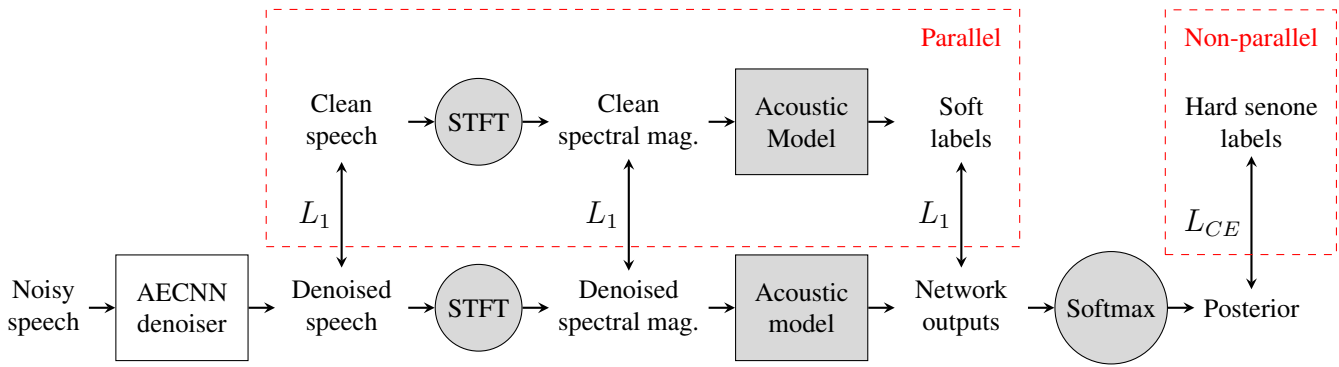

\section{Related Work}
\label{sec:related}

Speech enhancement is a rich field of work with a huge variety of techniques. Spectral feature based enhancement systems have focused on masking approaches \cite{virag1999single}, and have gained popularity with deep learning techniques \cite{narayanan2013ideal} for ideal ratio mask and ideal binary mask estimation \cite{wang2005ideal}.

\subsection{Perceptual Loss}

Perceptual losses are a form of knowledge transfer \cite{vapnik2015learning}, which is defined as the technique of adding auxiliary information at train time, to better inform the trained model. The first perceptual loss was introduced for the task of style transfer \cite{johnson2016perceptual}. These losses depends on a pre-trained network that can disentangle relevant factors. Two examples are fed through the network to generate a loss at a high level of the network. In style transfer, the perceptual loss ensures that the high-level contents of an image remain the same, while allowing the texture of the image to change.

For speech-related tasks a perceptual loss has been used to denoise time-domain speech data \cite{germain2018speech}, where the loss was called a "deep feature loss". The perceiving network was trained for acoustic environment detection and domestic audio tagging. The clean and denoised signals are both fed to this network, and a loss is computed at a higher level.

Perceptual loss has also been used for spectral-domain data, in the mimic loss framework. This has been used for spectral mapping for robust ASR in \cite{bagchi2018spectral} and \cite{plantinga2018exploration}. The perceiving network in this case is an acoustic model trained with senone targets. Clean and denoised spectral features are fed through the acoustic model, and a loss is computed from the outputs of the network. These works did not evaluate mimic loss for speech enhancement, nor did they develop the framework for use without parallel data.

\subsection{Enhancement Without Parallel Data}

One approach for enhancement without parallel data introduces an adversarial loss to generate realistic masks \cite{higuchi2017adversarial}. However, this work is only evaluated for ASR performance, and not speech enhancement performance.

For the related task of voice conversion, a sparse representation was used by  \cite{cisman2017sparse} to do conversion without parallel data. This wasn't evaluated on enhancement metrics or ASR metrics, but would prove an interesting approach.

Several recent works have investigated jointly training the acoustic model with a masking speech enhancement model~\cite{wang2016joint,menne2019investigation,soni2019label}, but these works did not evaluate their system on speech enhancement metrics. Indeed, our internal experiments show that without access to the clean data, joint training severely harms performance on these metrics.

\section{Mimic Loss for Enhancement}
\label{sec:mimic}

\begin{figure*}
    \centering
    \includegraphics[width=\linewidth]{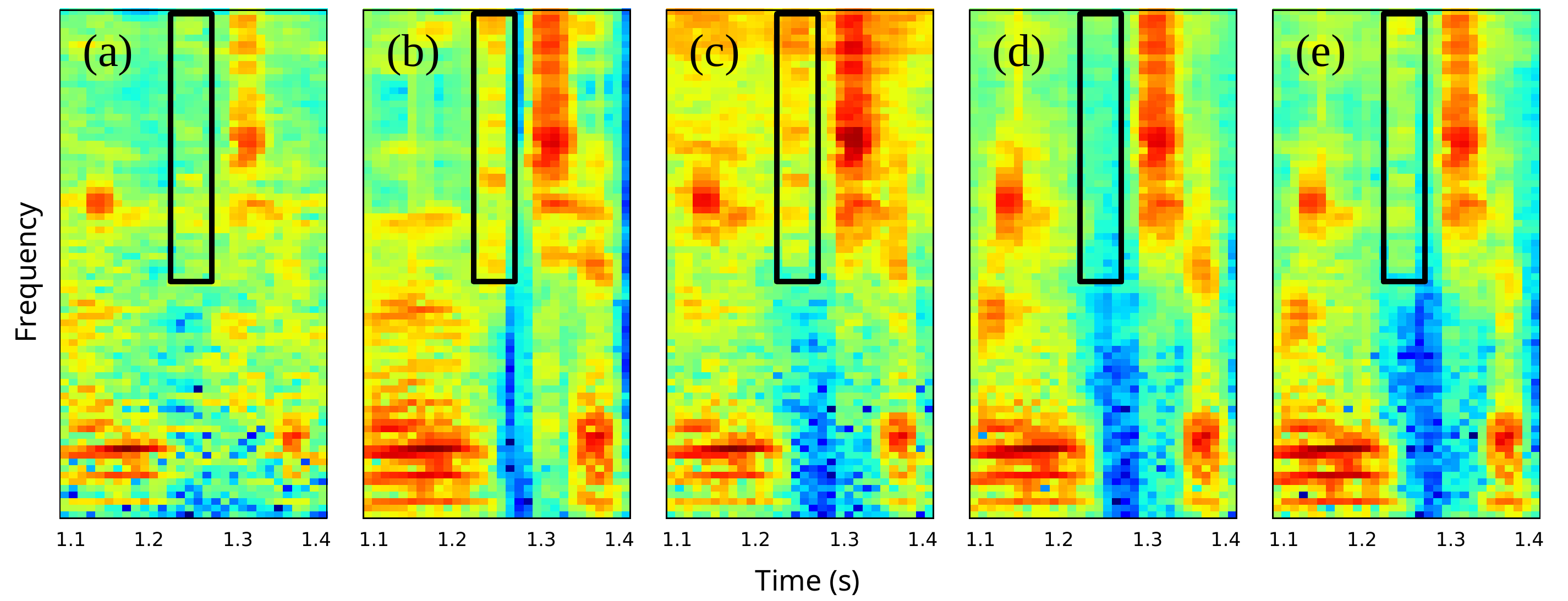}
    \caption{Comparison of a short segment of the log-mel filterbank features of utterance M06\_441C020F\_STR from the CHiME-4 corpus. The generation procedure for the features are as follows: (a)~noisy, (b)~clean, (c)~non-parallel mimic, (d)~local losses, (e)~local + mimic loss. Highlighted is a region enhanced by mimic loss but ignored by local losses.}
    \label{fig:compare}
\end{figure*}

As noted before, we build on the work by Pandey and Wang that denoises the speech signal in the time domain, but computes a mapping loss on the spectral magnitudes of the clean and denoised speech samples. This is possible because the STFT operation for computing the spectral features is fully differentiable. This framework for enhancement lends itself to other spectral processing techniques, such as mimic loss. 

In order to train this off-the-shelf denoiser using the mimic loss objective, we first train an acoustic model on clean spectral magnitudes. The training objective for this model is cross-entropy loss against hard senone targets. Crucially, the weights of the acoustic model are frozen during the training of the enhancement model. This prevents passing information from enhancement model to acoustic model in a manner other than by producing a signal that behaves like clean speech. This is in contrast to joint training, where the weights of the acoustic model are updated at the same time as the denoising model weights, which usually leads to a degradation in enhancement metrics.

Without parallel speech examples, we apply the mimic loss framework by using hard senone targets instead of soft targets. The loss against these hard targets is cross-entropy loss ($L_{CE}$). The senone labels can be gathered from a hard alignment of the transcripts with the noisy or denoised features; the process does not require clean speech samples. Since this method only has access to phone alignments and not clean spectra, we do not expect it to improve the speech quality, but expect it to improve intelligibility.

We also ran experiments on different formats for the mimic loss when parallel data is available. Setting the mapping losses to be $L_1$ was determined to be most effective by Pandey and Wang. For the mimic loss, we tried both teacher-student learning with $L_1$ and $L_2$ losses, and knowledge-distillation with various temperature parameters on the softmax outputs. We found that using $L_1$ loss on the pre-softmax outputs performed the best, likely due to the fact that the other losses are also $L_1$. When the loss types are different, one loss type usually comes to dominate, but each loss serves an important purpose here.

We provide an example of the effects of mimic loss, both with and without parallel data, by showing the log-mel filterbank features, seen in Figure \ref{fig:compare}. A set of relatively high-frequency and low-magnitude features is seen in the highlighted portion of the features. Since local metrics tend to emphasize regions of high energy differences, they miss this important phonetic information. However, in the mimic-loss-trained systems, this information is retained.

\section{Experiments}
\label{sec:experiments}

For all experiments, we use the CHiME-4 corpus, a popular corpus for robust ASR experiments, though it has not often been used for enhancement experiments. During training, we randomly select a channel for each example each epoch, and we evaluate our enhancement results on channel 5 of et05.

Before training the enhancement system, we train the acoustic model used for mimic loss on the clean spectral magnitudes available in CHiME-4. Our architecture is a Wide-ResNet-inspired model, that takes a whole utterance and produces a posterior over each frame. The model has 4 blocks of 3 layers, where the blocks have 128, 256, 512, 1024 filters respectively. The first layer of each block has a stride of 2, down-sampling the input. After the convolutional layers, the filters are divided into 16 parts, and each part is fed to a fully-connected layer, so the number of output posterior vectors is the same as the input frames. This is an utterance-level version of the model in~\cite{plantinga2018exploration}.

In the case of parallel data, the best results were obtained by training the network for only a few epochs (we used 5). However, when using hard targets, we achieved better results from using the fully-converged network. We suspect that the outputs of the converged network more closely reflect the one-hot nature of the senone labels, which makes training easier for the enhancement model when hard targets are used. On the other hand, only lightly training the acoustic model generates softer targets when parallel data is available.

For our enhancement model, we began with the state-of-the-art framework introduced by Pandey and Wang in \cite{pandey2019new}, called AECNN. We reproduce the architecture of their system, replacing the PReLU activations with leaky ReLU activations, since the performance is similar, but the leaky ReLU network has fewer parameters.

\subsection{Without parallel data}

We first train this network without the use of parallel data, using only the senone targets, and starting from random weights in the AECNN. In Table \ref{tab:unparallel} we see results for enhancement without parallel data: the cross-entropy loss with senone targets given a frozen clean-speech network is enough to improve eSTOI by 4.3 points.
This is a surprising improvement in intelligibility given the lack of parallel data, and demonstrates that phonetic information alone is powerful enough to provide improvements to speech intelligibility metrics.
The degradation in SI-SDR performance, a measure of speech quality, is expected, given that the denoising model does not have access to clean data, and may corrupt the phase.

We compare also against joint training of the enhancement model with the acoustic model. This is a common technique for robust ASR, but has not been evaluated for enhancement. With the hard targets, joint training performs poorly on enhancement, due to co-adaptation of the enhancement and acoustic model networks. 
Freezing the acoustic model network is critical since it requires the enhancement model to produce speech the acoustic model sees as ``clean.''

\begin{table}[]
    \centering
    \begin{tabular}{lccc}
        \toprule
        Features & SI-SDR & eSTOI \\
        \midrule
        Noisy speech & 7.5 & 68.3 \\
        Mimic - hard targets & 1.6 & 72.6 \\
        Joint training & 0.6 & 47.0 \\
        \bottomrule
    \end{tabular}
    \caption{Speech enhancement scores for the state-of-the-art architecture trained from scratch without the parallel clean speech data from the CHiME-4 corpus. Evaluation is done on channel 5 of the simulated et05 data. The joint training is done with an identical setup to the mimic system.}
    \label{tab:unparallel}
\end{table}

\subsection{With parallel data}

In addition to the setting without any parallel data, we show results given parallel data. In Table \ref{tab:parallel} we demonstrate that training the AECNN framework with mimic loss improves intelligibility over both the model trained with only time-domain loss (AECNN-T), as well as the model trained with both time-domain and spectral-domain losses (AECNN-T-SM).  We only see a small improvement in the SI-SDR, likely due to the fact that the mimic loss technique is designed to improve the recognizablity of the results. In fact, seeing any improvement in SI-SDR at all is a surprising result.

We also compare against joint training with an identical setup to the mimic setup (i.e. a combination of three losses: teacher-student loss against the clean outputs, spectral magnitude loss, and time-domain loss). The jointly trained acoustic model is initialized with the weights of the system trained on clean speech. We find that joint training performs much better on the enhancement metrics in this setup, though still not quite as well as the mimic setup. Compared to the previous experiment without parallel data, the presence of the spectral magnitude and time-domain losses likely keep the enhancement output more stable when joint training, at the cost of requiring parallel training data.

\begin{table}[]
    \centering
    \begin{tabular}{lccc}
        \toprule
        Features & SI-SDR & eSTOI \\
        \midrule
        Noisy speech & 7.5 & 68.3 \\
        \midrule
        AECNN-T & 11.5 & 77.0 \\
        + Mimic loss & 11.9 & 79.1 \\
        \midrule
        AECNN-T-SM & 11.7 & 78.9 \\
        + Mimic loss & 11.9 & 79.8 \\
        \midrule
        Joint training & 11.7 & 79.5 \\
        \bottomrule
    \end{tabular}
    \caption{Speech enhancement scores for the state-of-the-art system trained with the parallel data available in the CHiME-4 corpus. Evaluation is done on channel 5 of the simulation et05 data. Mimic loss is applied to the AECNN model trained with time-domain mapping loss only, as well as time-domain and spectral magnitude mapping losses. The joint training system is done with an identical setup to the mimic system with all three losses.}
    \label{tab:parallel}
\end{table}

\section{Conclusion}
\label{sec:conclusion}

We have shown that phonetic feedback is valuable for speech enhancement systems. In addition, we show that our approach to this feedback, the mimic loss framework, is useful in many scenarios: with and without the presence of parallel data, in both the enhancement and robust ASR scenarios. Using this framework, we show improvement on a state-of-the-art model for speech enhancement. The methodology is agnostic to the enhancement technique, so may be applicable to other differentiably trained enhancement modules.

In the future, we hope to address the reduction in speech quality scores when training without parallel data. One approach may be to add a GAN loss to the denoised time-domain signal, which may help with introduced distortions. In addition, we could soften the cross-entropy loss to an $L_1$ loss by generating "prototypical" posterior distributions for each senone, averaged across the training dataset.

Mimic loss as a framework allows for a rich space of future possibilities. To that end, we have made our code available at \url{http://github.com/OSU-slatelab/mimic-enhance}.

\bibliographystyle{IEEEbib}
\bibliography{refs}

\end{document}